\documentstyle[twoside,fleqn,espcrc2,epsfig]{article}

\newcommand{\AmS}{{\protect\the\textfont2
  A\kern-.1667em\lower.5ex\hbox{M}\kern-.125emS}}
\title{
{
\vspace{-4.5cm} \normalsize \hfill
\parbox{30mm}{ITP-Budapest 545\\
}
}\\[30mm]
Where does the  hot electroweak phase transition end?
      }

\author{F. Csikor, Z. Fodor \address{Institute for Theoretical Physics, E\"otv\"os University,
H-1088 Budapest, Hungary
 \\}
        and
       J. Heitger \address{Institut f\"ur Theoretische Physik I, Universit\"at M\"unster,
       D-48149 M\"unster, Germany}
	\thanks{Present address: DESY Zeuten,
	Platanenallee 6, D-15738 Zeuthen, Germany}
	}

\begin{document}
\begin{abstract}
We give the nonperturbative phase diagram of the four-dimensional
hot electroweak phase transition.
A systematic extrapolation  $a \rightarrow 0$
is done. Our results show that the finite temperature SU(2)-Higgs
phase transition is of first order for Higgs-boson masses
$m_H<66.5 \pm 1.4$ GeV.
The full four-dimensional result
agrees completely with that of the dimensional reduction
approximation.  This fact is of particular importance, because
it indicates that the fermionic sector of the Standard Model (SM) 
can be included perturbatively.  We obtain that the Higgs-boson endpoint
mass in the SM is  $72.4 \pm 1.7$ GeV. Taking into account
the LEP Higgs-boson mass lower bound excludes
any electroweak phase transition in the SM.
\end{abstract}

\maketitle


The observed baryon asymmetry is finally determined at the
electroweak phase transition (EWPT) \cite{KuRS}.
The perturbative approach breaks down for the physically
allowed Higgs-boson masses (e.g. $m_H>70$ GeV) \cite{pert}.
Since merely the bosonic sector is
responsible for the bad perturbative features (due to infrared problems)
the simulations are done without the inclusion of fermions 
on four-dimensional lattices \cite{4d}, \cite{4d-rev}. Another 
approach is the 
simulations of the
reduced model in three-dimensions 
 \cite{3d}, \cite{3d-rev}. 
 The comparison of the results
obtained by the two techniques is not only a useful
cross-check on the perturbative reduction procedure
but also a necessity  because the fermions 
must be included perturbatively.

The line 
 separating the symmetric and broken phases
on the $m_H-T_c$ plane has an endpoint, $m_{H,c}$. In 
four dimension at $m_H \approx 80$ GeV the EWPT turned out
to be extremely weak, even consistent with the no phase
transition scenario on the 1.5-$\sigma$ level \cite{4d80}.
Three-dimensional results show that for $m_H>95$ GeV
no first order phase transition exists \cite{3d95} and more specifically
that the endpoint is  $m_{H,c} \approx 67$ GeV \cite{3d80}. In this
paper  we present the analysis of the endpoint on four
dimensional lattices in the continuum limit and transform the result to the full SM.


We will use 
 different spacings in temporal ($a_t$) and spatial
($a_s$) directions. The asymmetry of the
lattice spacings is given by the asymmetry factor $\xi=a_s/a_t$.
The action reads
\begin{eqnarray*}
 S[U,\varphi]= \;\;\;\;\;\;\;\;\;\;\;\;\;\;\;\;\;\;\;\;\;\;\;\;\;\;\;\;\;
 \;\;\;\;\;\;\;\;\;\;\;\;\;\;\;\;\;\;\;\;\;\;\;\;\;\; \\
 \beta_s \sum_{sp}
\left( 1 - {1 \over 2} {\rm Tr\,} U_{pl} \right)
+\beta_t \sum_{tp}
\left( 1 - {1 \over 2} {\rm Tr\,} U_{pl} \right)
 \\
\left.
-\kappa_s\sum_{\mu=1}^3
{\rm Tr\,}(\varphi^+_{x+\hat{\mu}}U_{x,\mu}\,\varphi_x)
-\kappa_t {\rm Tr\,}(\varphi^+_{x+\hat{4}}U_{x,4}\,\varphi_x)
\right.  \\
+ \sum_x \left\{ {1 \over 2}{\rm Tr\,}(\varphi_x^+\varphi_x)+
\lambda \left[ {1 \over 2}{\rm Tr\,}(\varphi_x^+\varphi_x) - 1 \right]^2
\right\},
\end{eqnarray*}
where $U_{x,\mu}$ denotes the SU(2) gauge link variable,  $U_{sp}$ and
$U_{tp}$
the path-ordered product of the four $U_{x,\mu}$ around a
space-space or space-time plaquette, respectively;
$\varphi_x$ stands for the Higgs field. The anisotropies
$\gamma_\beta^2=\beta_t/\beta_s$ and $\gamma_\kappa^2=\kappa_t/\kappa_s$
are functions of the asymmetry $\xi$. These functions have been
determined perturbatively \cite{T0pert} and non-perturbatively
\cite{T0nonpert}.
 In this paper we use the asymmetry parameter
$\xi=4.052$, which gives $\gamma_\kappa=4$ and $\gamma_\beta=3.919$.

We have performed our simulations on finer and finer lattices,
moving along the lines of constant physics (LCP). In our case
there are three bare parameters ($\kappa, \beta, \lambda $). The bare
parameters are chosen in a way that the zero temperature renormalized gauge
coupling $g_R$ is held constant and the mass ratio for the Higgs- and W-bosons
$R_{HW}=m_H/m_W$ corresponds to the Higgs mass at the endpoint of
first order phase transitions: $R_{HW,c}$. These two conditions determine
a LCP as
a one-dimensional subspace in the original space of bare parameters.
The position on the LCP gives the lattice spacing $a$. As the lattice
spacing decreases $R_{HW,c} \rightarrow R_{HW,cont.}$.

Since our theory is a bosonic one we assumed that the corrections
are quadratic in the lattice spacings; thus an $a^2 $ fit has been
performed for  $R_{HW,c} $ to determine its continuum value.

We have carried out $T \neq 0$
simulations  on $L_t= 2,3,4,5$ lattices 
 and tuned $\kappa $ to the transition
point.  This condition fixes the lattice spacings: $a_t =a_s / \xi \, = \,
1/(T_c L_t ) $ in terms of the transition temperature $T_c $ in physical units.
The third parameter $\lambda $, finally specifying the physical Higgs mass
in lattice units, has been chosen   in a way that the transition
corresponds to the endpoint of the first order phase transition subspace.

In this paper $V=L_t\cdot L_s^3$ type four-dimensional
lattices are used. For each $L_t$ we had 8 different lattices, each
of them had approximately twice as large lattice-volume as the previous one.
The smallest lattice was $V=2\cdot 5^3$ and the largest one
was $V=5\cdot 50^3$. 
Reweighting was used to obtain information
in the vicinity of a simulation point.

The determination of the endpoint of the finite temperature
EWPT is done by the use of the Lee-Yang zeros of the 
partition function ${\cal Z}$ \cite{LY}.
Near the first order phase transition point the partition function reads
${\cal Z}={\cal Z}_s + {\cal Z}_b \propto \exp (-V f_s ) + \exp ( -V f_b ) \, ,
$ where the indices s(b) refer to the symmetric (broken) phase and $f$ stands
for the free-energy densities. Near the phase transition point we also have
$f_b = f_s + \alpha (\kappa - \kappa _c ) \, ,$
which shows that for complex $\kappa$ values ${\cal Z}$ vanishes at
$  {\rm Im} (\kappa )= \pi \cdot (n-1/2) / (V\alpha )$
 for integer $n$.  In case a first order phase transition is present,
these Lee-Yang
zeros move to the real axis as the volume goes to infinity. In case a
phase transition is absent the Lee-Yang
zeros stay away from the real $\kappa $ axis. 
Denoting
$\kappa_0$ the lowest zero of ${\cal Z}$, i.e. the  position of the zero
closest zero to the real axis, one expects in the vicinity of
the endpoint the scaling law
${\rm Im}(\kappa_0)=c_1(L_t,\lambda)V^\nu+c_2(L_t,\lambda)$.
In order to pin down the endpoint we are looking
for a $\lambda$ value for which $c_2$ vanish.
In practice we analytically continue ${\cal Z}$ to complex
values of $\kappa $ by reweighting the available data.
Also small changes  in
$\lambda$ have been done by reweighting.
As an example, the dependence of $c_2$ on  $\lambda$ for $L_t =3$
is shown in fig. 2. To determine the critical value of $\lambda$
i.e. the largest value, where $c_2=0$, we
have performed  fits linear in $\lambda$ to the nonnegative $c_2$ values.

\begin{figure}[htb]
\vspace{-0.5cm}
\centerline{\epsfxsize=0.65 \linewidth \epsfbox{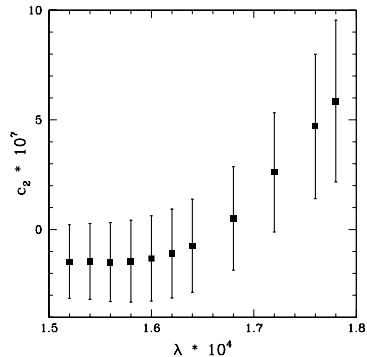}}
\vspace{-0.9cm}
\caption{
Dependence of $c_2$ on $\lambda$ for $L_t=3$.
}
\label{fig2}
\end{figure}

Having determined the endpoint
$\lambda_{crit.} (L_t)$ for each $L_t$ we calculate the $T=0$
quantities ($R_{HW},g_R^2$) on $V=(32L_t)\cdot (8L_t)\cdot (6L_t)^3$ lattices,
where $32L_t$ belongs to the temporal extension, and
extrapolate to the continuum limit. 
Having established the correspondence
between $\lambda_{crit.} (L_t)$ and $R_{HW}$, the $L_t$ dependence of the
critical $R_{HW}$ is easily obtained. Fig. 3 shows the dependence of
the endpoint $R_{HW}$ values on $1/L_t^2 $. A linear extrapolation
in $1/L_t^2 $ yields the continuum limit
value of the endpoint
$R_{HW}$. We obtain $66.5 \pm 1.4$ GeV, which is our final result.

\begin{figure}[htb]
\vspace{-0.5cm}
\centerline{\epsfxsize=0.65 \linewidth \epsfbox{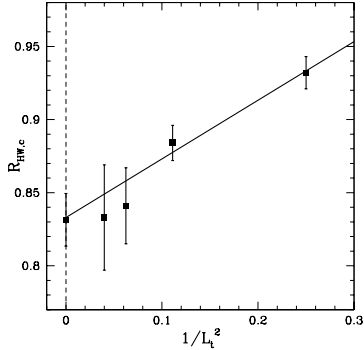}}
\vspace{-0.9cm}
\caption{
Dependence of $R_{HW,cr}$, i.e. $R_{HW}$ corresponding to the endpoint
of first order phase transitions on $1/L_t^2$ and extrapolation to the
infinite volume limit.
 }
  \label{fig3}
   \end{figure}

Comparing our result to those of the 3d analyses \cite{3d80} one
observes complete
agreement. Since the error bars on the endpoint determinations are on the
few percent level, the uncertainty of the dimensional reduction procedure
is also in this range. This indicates that the analogous perturbative
inclusion  of the fermionic sector results also in few percent error on
$M_H$.

Based on our published data \cite{4d-rev,T0nonpert} and the present results 
 we are able to draw the phase diagram of the
SU(2)-Higgs model in the ($T_c /m_H - R_{HW} $) plane. This is shown in
fig. 4. The continuous line -- representing the phase-boundary -- is
a quadratic fit to the data points.

\begin{figure}[htb]
\vspace{-0.5cm}
\centerline{\epsfxsize=0.65 \linewidth \epsfbox{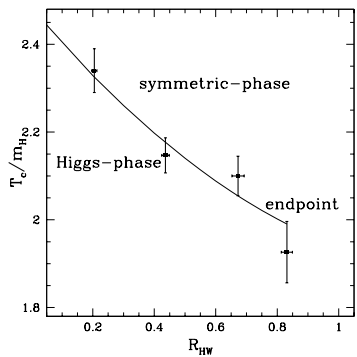}}
\vspace{-0.9cm}
\caption{
Phase diagram of the SU(2)-Higgs model in the ($T_c /m_H - R_{HW} $)
plane.
}
\label{fig3v}
\end{figure}

Finally, we determine what is the endpoint value in the full SM.
Our nonperturbative analysis shows that the perturbative integration of
the heavy modes is correct within our error bars. Therefore we use perturbation
theory \cite{KLRS96} to transform the SU(2)-Higgs model endpoint value to
the full SM. We obtain $72.4 \pm 1.7$ GeV, where the error
includes the measured error of $R_{HW,cont.}$, $g_R^2$ and the estimated
uncertainty \cite{Laine96}
due to the different definitions of the
gauge couplings between this paper and \cite{KLRS96}. The dominant error
comes from the uncertainty on the position of the endpoint.

  The present experimental lower limit of the SM Higgs-boson mass
is $89.8$ GeV \cite{LEP}. Taking into account all errors (in particular
those coming from integrating out the heavy fermionic modes),
our endpoint value excludes the
possibility of any EWPT in the SM.
This also means that the SM baryogenesis in the early Universe
is ruled out.

For details of this analysis  see \cite{tobe}.

Simulations have been carried out on the Cray-T90 at HLRZ-J\"ulich, 
on the APE-Quadrics at DESY-Zeuthen and on the PMS-8G PC-farm
in Budapest. This work was supported by 
 OTKA-T016240/T022929 and FKP-0128/1997.

\end{document}